\documentclass[11pt,letterpaper]{article}
\usepackage[utf8]{inputenc}
\usepackage[T1]{fontenc}
\usepackage[english]{babel}
\usepackage{amsmath}
\usepackage{amsfonts}
\usepackage{amssymb}
\usepackage{graphicx}
\usepackage[margin=0.97in]{geometry}
\author{Francisco A. Dom\'inguez-Serna$^{1,*}$,Fernando Rojas$^{2}$,Karina Garay-Palmett$^{3,\dagger}$\\ 
$^{1,*}$ C\'atedras Conacyt - Centro de Investigaci\'on Cient\'ifica y de Educaci\'on Superior de \\Ensenada, B.C., 22860, M\'exico\\
$^{2}$ Centro de Nanociencias y Nanotecnolog\'ia, Departamento de Fisica,\\ Universidad Nacional Aut\'onoma de M\'exico, Ensenada, B. C., 22860, Mexico\\
$^{3}$ Departamento de \'Optica, Centro de Investigaci\'on Cient\'ifica y de Educaci\'on Superior \\ de Ensenada, B.C., 22860, M\'exico\\
}

\usepackage{enumitem}
\def\ket#1{\mathinner{|{#1}\rangle}}
\def\bra#1{\mathinner{\langle{#1}|}}
\def\KetC#1{\left|\left\{ #1 \right\} \right\rangle}
\def\BraC#1{\left\langle \left\{ #1 \right\} \right|}

\newcommand\blfootnote[1]{%
	\begingroup
	\renewcommand\thefootnote{}\footnote{#1}%
	\addtocounter{footnote}{-1}%
	\endgroup
}

\title{Quantum teleportation with hybrid entangled resources prepared from heralded quantum states}

\begin{document}
	\maketitle
	
	\blfootnote{*fadomin@cicese.mx} \\
	\blfootnote{${}^\dagger$ kgaray@cicese.mx}
	\blfootnote{Final version of this manuscript: https://doi.org/10.1364/JOSAB.377687}
	
		© 2020 Optical Society of America. One print or electronic copy may be made for personal use only. Systematic reproduction and distribution, duplication of any material in this paper for a fee or for commercial purposes, or modifications of the content of this paper are prohibited.

	\begin{abstract}
		In this work we propose the generation of a hybrid entangled resource (HER) and its further application in a quantum teleportation scheme from an experimentally feasible point of view. The source for HER preparation is based on the four wave mixing process in a photonic crystal fiber, from which one party of its output bipartite state is used to herald a single photon or a single photon added coherent state. From the heralded state and linear optics the HER is created. In the proposed teleportation protocol Bob uses the HER to teleport qubits with different spectral properties. Bob makes a Bell measurement in the single photon basis and characterizes the scheme with its average quantum teleportation fidelity. Fidelities close to one are expected for qubits in a wide spectral range. The work also includes a discussion about the fidelity dependence on the geometrical properties of the medium through which the HER is generated. An important remark is that no spectral filtering is employed in the heralding process, which emphasizes the feasibility of this scheme without compromising photon flux. 
	\end{abstract}

	\section{Introduction}
	
	Quantum entanglement has become a key resource to perform quantum information tasks \cite{Bennett2000}. Most quantum communication protocols rely on the usage of a maximally entangled state shared among parties that want to exchange information \cite{Horodecki2009,Sergienko2005,Prevedel2007}. Quantum teleportation has also been identified as a valuable resource to perform quantum information operations, secret sharing applications, and measurement-based quantum information processing  \cite{Gottesman1999,Nielsen1998}.
	
	Quantum teleportation (QT) protocols, as initially proposed by Bennett, Brassard and Cr\'epeau \cite{Bennettetal-}, are established in a discrete basis, where a Bell state is shared between Alice (the sender) and Bob (the receiver), and in which Alice wants to teleport the state of a particle to Bob. To fulfil this purpose, Alice performs a joint measurement on her part of the shared entangled state and the state to be teleported, in the Bell basis. With this procedure, the state initially possessed by Alice is teleported to Bob up to a unitary transformation, which is determined by the outcome of the Bell measurement. In the ideal scenario, the state is perfectly teleported from Alice to Bob.
	
	The experimental implementation of QT came promptly in the discrete variable (DV) basis as demonstrated first in ref. \cite{Bouwmeester97a}, where the polarization state of a photon was teleported, via a maximally entangled state obtained by parametric downconversion in a nonlinear crystal. The experimental demonstration of QT in matter came later, supported by nuclear magnetic resonance \cite{Nielsen1998}, which makes QT feasible at inter-atomic distances. QT has been implemented in many other different schemes and degrees of freedom, as shown in ref. \cite{Pirandola2015} (and references therein).
	
	It is well know that photons are low sensitive to decoherence, thereby they are regarded among the best for  long distance implementations of quantum protocols \cite{Jayakumar2014,Obrien2010,OBrien2007}. However, discrete variable proposals with single photons lack of full Bell basis distinguishability, which has been overcome with the aid of continuous variable (CV) protocols \cite{Furusawa1998}. Unfortunately, maximally entangled states cannot be prepared for these schemes, given the impossibility to generate infinitely squeezed states \cite{Pirandola2015,Andersen2013a}, and therefore, high fidelities are difficult to obtain. The hybrid DV-CV approach, initially studied in ref. \cite{Andersen2013a} and further developed in refs. \cite{Takeda2013b,Marshall2014}, exploits the best of both domains, with the entangled resource in one domain and the state to be teleported in the other one. A different alternative is found with hybrid entangled resources, where the maximally entangled state between two subsystems is formed by a superposition of CV and DV domains. For such hybrid entangled state (HES), a projective measure in the discrete or continuous basis on a subsystem will project the another one to the opposite\textit{ DV-CV domain \cite{Nogueira2013,Jeong2014}}. 
	
	Hybrid entangled states have also been studied as a resource for quantum teleportation. Hybrid QT schemes are an alternative to teleport information contained in discrete or continuous degrees of freedom of a system to different degrees of freedom (discrete or continuous variable) of other systems. In ref.~\cite{Wen-Yuan2007}, quantum teleportation of linear combinations of atomic states is achieved with the aid of a HES formed by coherent states and atomic states. The preparation of a HES between CV and DV at remote places connected by a lossy channel was first demonstrated by Morin \textit{et. al.} \cite{Morin2014}, with its further application to teleport a CV state and convert it into a DV one \cite{Ulanov2017}. HES formed by combinations of coherent states (CS) and Fock states have been theoretically and experimentally studied, as shown in refs. \cite{Sekatski2012b, Jeong2014}, in which the advantages of these states for phase detection and its robustness against scattering are pointed out by the authors. Following this direction,  we previously studied the non-classical properties of HES resulting from the linear combination of single-photon-added coherent states (SPACS) and coherent states, under the presence of noise \cite{Dominguez-Serna2017}. Photon-added-coherent states are obtained by the successive  application of the creation operator over a CS \cite{Agarwal1991}. It is remarkable that beyond QT proposals involving hybrid entangled resources, to our knowledge no schemes of quantum teleportation based on broadband states, other than squeezed, have been reported \cite{Benichi2011,VanLoock2000,Yonezawa2007,Pirandola2015}.
	
	In this study we propose a hybrid quantum teleportation scheme in continuity to our preliminary study in ref. \cite{Dominguez-Serna2017a}, in which all involved parties are let to be broadband and can exhibit different spectral properties. We also include simulation results for a specific and experimentally feasible situation that will serve as a figure of merit for potential applications of our proposal. At the same time, this work is motivated in the fewer resources that path qubit and entanglement needs on the long distance communication architectures \cite{Monteiro2015}. 
	
	This paper is organized as follows: in section two we propose the HER generation from a third order non-linear process in a commercial photonic crystal fiber (PCF) and detail the optimal parameters found for the resource. In section three, the QT protocol is outlined and an averaged version of the fidelity is analyzed in order to characterize the potential applicability of this proposal. Finally, conclusions and comments on future work are included.
	
	\section{Hybrid entangled resource preparation and quantum teleportation protocol}

\begin{figure}[ht!]
		\centering
		\includegraphics[width=8.5cm]{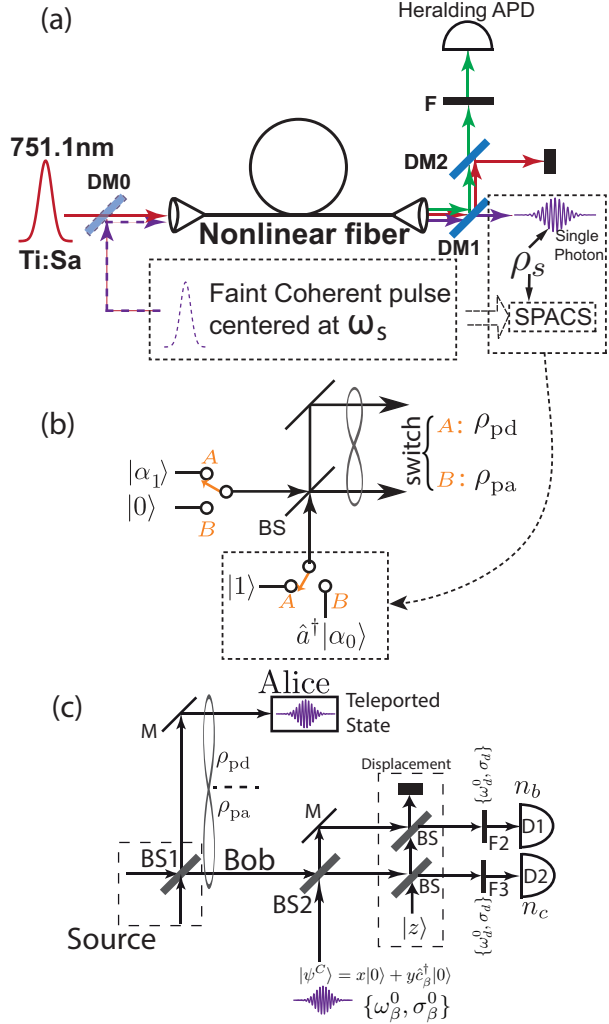}
		\caption{(a) Scheme for the heralding process of a SPS or SPACS generated by FWM, and given by Eq.~(\ref{eq:heralded}). Detection of an idler single-photon in the APD heralds a SPS or a SPACS, depending on the faint coherent pulse is absent or present, respectively. In the figure, DM is for dichroic mirror, and F for spectral filtering. (b) Schematic representation of the HER preparation by combining vacuum or SPS with a CS or SPACS in a BS depending on the switches positions. (c) Scheme for the QT protocol, where $\rho_{pd(pa)}$ is the HER as described in Eq. (\ref{eq:rhoPAPDSFWM}), the qubit to be teleported from Bob to Alice is $\ket{\psi^C}$ is mixed with the HER, and a joint measurement is accomplished by the displacement section and photodetection.}
		\label{fig:Esquema}
	\end{figure}

	In this section we address the hybrid entangled resource preparation and its further application on a quantum teleportation protocol. We propose the HER generation through the four wave mixing process, which is a third-order nonlinear interaction widely exploited for photon pair generation in optical fiber and waveguide \cite{Chen2005,Takesue2012,Suhara2009,Dudley2009}. The schemes for HER generation and the QT protocol are better explained with the aid of Fig. \ref{fig:Esquema}. Panel (a) outlines the generation of the needed sources to produce the HER; panel (b) is a schematic representation of the HER preparation where the switches positions determine the generation of two different HER by selecting two pairs of states to be combined in a beamsplitter  resource generation (process discussed in section \ref{HERproposal}); while panel (c) depicts the QT protocol that will be further analyzed in section \ref{QTproposal}.

	\subsection{Resource preparation}\label{HERproposal}
	The hybrid entangled resource is proposed to be generated by the combination in a beamsplitter (BS) of a single-photon state (SPS) or a single-photon-added coherent state (SPACS), with the vacuum state or a coherent state. The proposal addresses the generation of SPS's and SPACS's by means of the FWM process in an optical fiber. Both types of states can be prepared in the same experimental setup with the only difference that for the former the nonlinear process must be in the spontaneous regime i.e., mediated by vacuum fluctuations; and for the later, the initial state is populated with a weak coherent state allowing for a SPACS generation in the corresponding output mode. This process can be viewed as the generation of a state of the form $\ket{ \varphi_0 } = \mathcal{N} \hat{a}^\dagger_s \ket{\alpha_0}_s \ket{1}_i$, which after detection of a photon in the $i$-mode is projected to a state that can be a SPACS for $\alpha_0 \neq 0$, or a SPS for $\alpha_0 =0$. Here, $\ket{\alpha_0}$ represents a coherent state. For HER generation the heralded-SPACS (or heralded-SPS) is combined in a BS with another CS ($\ket{\alpha_1}$), which can lead to the following two different cases: i) the heralded-state resulting from $\alpha_0 \neq 0$ is combined in the BS with the vacuum state (i.e. $\alpha_1 = 0$), and ii) the heralded-state for $\alpha_0 = 0$ is combined in the BS with the CS characterized by $\alpha_1 \neq 0$.
	
	In the ideal scenario, the output state of the BS acquires the following form $
	\ket{\varphi_1}_{pd(pa)} = \mathcal{N} (\hat{a}^\dagger_s \ket{\alpha}_s\ket{\alpha}_i \mp  \hat{b}^\dagger_i\ket{\alpha}_s\ket{\alpha}_i  )
	$. Note that $pd(pa)$ subscripts were added, as we will refer to these superpositions as "photon displaced" and "photon added", where the photon displaced tag makes sense when the unitary transformation $(\hat{D}_s(-\alpha)\hat{D}_i(-\alpha))$ is applied to $\ket{\varphi_1}$. In general $\alpha$ is function of $\alpha_0$ and the beamsplitter parameters.
	
	Now we proceed to model the generation of the HER in a more realistic situation. First, a pump pulse is sent to a nonlinear fiber phase-matched to the generation of frequency non-degenerate photon pairs, as shown in Fig. \ref{fig:Esquema}(a), where purple and green lines represent signal and idler fields, respectively. We consider a co-linear and co-polarized FWM configuration for the sake of simplicity, interaction that can be described by the following Hamiltonian  \cite{Rohde2007,URen2005,Mosley2007},
	\begin{equation}
	\hat{H}(t)=\frac{3}{4}\epsilon_0 \chi^{(3)}\!\! \int \!\!d^3 \mathbf{r}   \hat{E}^{(+)}_1(\mathbf{r},t) \hat{E}^{(+)}_2(\mathbf{r},t) \hat{E}^{(- )}_s(\mathbf{r},t) \hat{E}^{(-)}_i(\mathbf{r},t) + \text{h.c.},
	\label{eq:Hamiltonian}
	\end{equation}
	\noindent where $\chi^{(3)}$ is the third-order nonlinear susceptibility of the medium, $\epsilon_0$ is the vacuum electric permittivity and $\hat{E}_j$ the $j$-th involved optical electric field. 
	
	In this paper, we study the generation of a single photon added coherent state in the signal mode and a single photon in the idler mode. This can be accomplished by injecting a CS to the nonlinear medium spatio-temporally matched with the corresponding signal mode in eq. (\ref{eq:Hamiltonian}). This CS is depicted in Fig. \ref{fig:Esquema}(a) as a purple dashed-arrow related to a faint coherent pulse centered at $\omega_s$. The effect of this can be obtained by a standard perturbative approach to the first order, in which the initial states are $\KetC{\alpha_0}_s\ket{0}_i$. This results in the following two-photon like quantum state
	\begin{equation}
	\begin{aligned}
	\ket{\Psi_2} &=\KetC{\alpha_0}_s\ket{0}_i\\ 
	&\quad+ \kappa\!\!\int\!\!\int\! d\omega_s d\omega_i F(\omega_s,\omega_i) \hat{a}^\dagger(\omega_s)\KetC{\alpha_0}_s\hat{a}^\dagger (\omega_i) \ket{0}_i,
	\end{aligned}
	\label{eq:JSAtotalPA}
	\end{equation}
	for which $\KetC{\alpha_0}_s$ is the seeding coherent state in a spatio-temporal mode that matches (at least partially) the signal, and that could have different spectral properties. If $\alpha_0=0$ this is the two photon state generated by spontaneous four wave mixing (SFWM). $\hat{a}^\dagger(\omega_{s}) (\hat{a}^\dagger(\omega_{i}))$ is the creation operator of a photon of frequency $\omega_s (\omega_i)$. In eq.~(\ref{eq:JSAtotalPA}) the joint function $F(\omega_s,\omega_i) $ contains the spectral correlation information of the generated state, which exhibits a high dependence on the geometrical parameters of the fiber and the pump spectral profiles \cite{Garay-Palmett2007}. Note that in this treatment we let $\KetC{\alpha_0}_s$ be distinct from vacuum to include a more general description. 
	
	For the heralding process, a set of dichroic mirrors, DM1 and DM2, are proposed to be used for separating the fields corresponding to signal and idler modes, and also a spectral filter F centered at frequency $\omega_i$ with bandwidth $\sigma_i$ located in front of an avalanche photo-diode (APD in Fig. \ref{fig:Esquema}(a)). The filter is included to maintain a general treatment; nevertheless, we will assume later, in a specific situation analyzed that this filter is not present. Note that a detection of an idler photon by the APD heralds the existence of a signal state, that could be a single photon state or a SPACS, in the form $\rho_{s}=\text{Tr}_i \left[  \hat{\Pi}_i\otimes \mathbb{I}_s \ket{\Psi_2}\bra{\Psi_2}  \hat{\Pi}_i\otimes \mathbb{I}_s  /\text{Tr}[\hat{\Pi}_i\otimes \mathbb{I}_s \ket{\Psi_2}\bra{\Psi_2}] \right]$, where the the operator $\hat{\Pi}_i\equiv \hat{\Pi}_i(n=1)$ stands for the detection of a single photon in the idler mode of the state in eq. (\ref{eq:JSAtotalPA}), and $\rho_{s}$ is the density matrix of the subsystem $s$, note that density matrices are written as bare operators throughout the manuscript $\rho \equiv \hat{\rho}$. In general, the multiple photon heralding process can be described with the positive operator valued measure $\hat{\Pi}_{i} (n)=\int d\omega_i \tilde{d}(\omega_i) \ket{n(\omega_i)} \bra{n(\omega_i)}$, where $\ket{n(\omega_i)}$ is a Fock state of frequency $\omega_i$ and $\tilde{d}(\omega_i)$ contains the spectral characteristics of the detector.  Here, it is also assumed that the probability of multiphoton generation is negligible, assumption which remains valid for small emitted mean photon number \cite{Mandel1995,Goldschmidt2008}. Therefore, the non-normalized heralded SPACS (or heralded SPS) is characterized by the following reduced density matrix operator
	
	\begin{equation}
	\tilde{\rho}_s=\int d\omega_s' d\omega_s'' \mathcal{G} (\omega_s',\omega_s'') \hat{a}^\dagger (\omega_s') \ket{\{ \alpha_0\}}\bra{\{ \alpha_0\}}\hat{a}(\omega_s''),
	\label{eq:heralded}
	\end{equation}
	
	\noindent with $\alpha_0\neq0$ (or $\alpha_0=0$). The convention of a tilde for non-normalized operators $\tilde{\cdot}$ is used throughout the manuscript. Here, it has been defined the function $\mathcal{G} (\omega_s',\omega_s'')$, given in terms of the joint function $F(\omega_s,\omega_i)$ appearing in eq. (\ref{eq:JSAtotalPA}) as
	
	\begin{equation}
	\mathcal{G} (\omega_s',\omega_s'') = \int d\omega_i F(\omega_s',\omega_i) F^*(\omega_s'',\omega_i)|\tilde{d}(\omega_i)|^2.
	\label{eq:edoEntangledGen}
	\end{equation}
	
	Continuing with the procedure for HER preparation, the heralded state in eq.~(\ref{eq:heralded}) is combined in a beamspliter with another CS $\KetC{\alpha_1}$ . This process is described by the action of the beamsplitter operator, $\hat{U}_{BS}$, in the following form
	\begin{equation}
	\rho_{pd(pa)}= \hat{U}_{BS} \rho^0_{pd(pa)} \hat{U}_{BS}^\dagger,
	\end{equation}
	where 
	\begin{equation}
	\rho^0_{pa} = \rho_s^{\alpha_0\neq 0} \otimes \ket{ 0  }\bra{ 0  }, \quad \rho^0_{pd} = \rho_s^{\alpha_0 = 0}\otimes \ket{\{\alpha_1 \}  }\bra{ \{\alpha_1 \} }.
	\label{eq:rhocero}
	\end{equation}
	The notation $ \rho_s^{\alpha_0\neq 0} ( \rho_s^{\alpha_0 = 0})$ has been used to emphasize the nature of the heralded state $\rho_s$ that depends on the value of $\alpha_0$ as shown in Eq. (\ref{eq:JSAtotalPA}). Emphasizing, $\rho_s$ is required as input to the BS together with the vacuum state or a CS to produce one HER or the other. The nomenclature $pd(pa)$ is motivated by the fact that $\rho_{pd}$ can be written equivalently as a superposition of single photon displaced states, while $\rho_{pa}$ can only be written as a superposition of photon-added coherent states. Here, $\{ \alpha_1\}$ represents a CS with spectral properties different to those of $\rho_s$. Despite the fact that we are considering the spatio-temporal properties of each of the states combined in the BS as closely related to each other, they are not assumed identical in general, i.e., we take into account the likely imperfect overlap present in experimental conditions. Without loss of generality, in this treatment we consider that bandwidths are different for each of the interacting fields. 
	
	By substitution of the heralded density matrix (Eq. (\ref{eq:heralded})), it is shown that Eq. (\ref{eq:edoEntangledGen}) can be written up to a normalization constant as follows:
	
	\begin{align}
	\label{eq:rhoPAPDSFWM}
	\tilde{\rho}_{pd(pa)}^{AB} &=\int d\omega_s' d\omega_s'' \mathcal{G} (\omega_s',\omega_s'') [ \hat{a}^{\dagger}(\omega_s')\mp \hat{b}^{\dagger}(\omega_s') ]\\ \nonumber &\times
	\KetC{ \alpha}\KetC{ \alpha} \BraC{ \alpha}\BraC{ \alpha} [ \hat{a}(\omega_s'')\mp \hat{b}(\omega_s'')],
	\end{align}
	
	\noindent where we defined  $\alpha:=\alpha_0/\sqrt{2}=\alpha_1/\sqrt{2}$. The entanglement properties of a state in the form given as in Eq. (\ref{eq:rhoPAPDSFWM}) have already been studied by our group and, we found that the negative superposition is more robust to decoherence  \cite{Dominguez-Serna2017}. For the sake of completeness we will include in the present study results for both superpositions. The HER described in Eq. (\ref{eq:rhoPAPDSFWM}) resembles a CV path entanglement state, which are known for requiring fewer resources than other proposals \cite{Monteiro2015}. The characterization of the HER used as a channel can be made, for instance, by means of simultaneous Homodyne detection of both arms of $\rho_{pd(pa)}^{AB}$ so as to obtain the nonclassicality of the resource \cite{Dominguez-Serna2017}, without the need of any additional displacement operation to perform Bell-like measurements.

	\subsection{Quantum teleportation protocol}\label{QTproposal}
	Now, we put forward a quantum teleportation protocol based on the HER detailed in previous section. We propose to use this state as a shared entangled channel between Bob (locally) and Alice (remotely) to establish the QT protocol. Figure \ref{fig:Esquema}(c) shows the proposed teleportation scheme: BS1 serves to generate the HER as described in previous section with aid of Fig \ref{fig:Esquema}(b), where the BS1 inputs depend on the HER to be created, $\rho_{\text{pa}}^{AB}$ or $\rho_{\text{pd}}^{AB}$. The steps of the QT protocol are described as follows:
	
	\begin{enumerate}[label=(\roman*)]
		\item Alice and Bob share the HER state, created as in the previous section, see Eq. (\ref{eq:rhoPAPDSFWM}).
		\item Bob wants to teleport the state 
		\begin{align}
		\label{psiC}
		\ket{\psi^C}= x\ket{0} + y \hat{c}^\dagger_\beta \ket{0},
		\end{align}
		
		\noindent where $\hat{c}^\dagger_\beta$ creates a  Gaussian  single-photon wave-packet with central wavelength $\lambda_\beta$ and bandwidth $\sigma_\beta$.
		
		\item Bob mixes his qubit (given by Eq. (\ref{psiC})) with his part of the HER state, through a second beamsplitter BS2, process which is described by the following operation
		\begin{align}
		\rho^{ABC}_{pd(pa)} =  \hat{U}_{BS} \rho^C \otimes \rho_{pd(pa)}^{AB}  \hat{U}_{BS}^\dagger,
		\end{align}
		where $\rho_C$ is the density matrix of the state in eq. (\ref{psiC})), $\rho^C = \ket{\psi^C} \bra{\psi^C}$.
		\item\label{numbasis} Bob performs a joint measurement in the displaced number basis, this measurement is restricted to $\alpha/2=iz\sqrt{1-\tau}$, where $z$ is the amplitude of the coherent state $\ket{z}$, as shown in Fig. \ref{fig:Esquema}(c), and $\tau$ is the transmittance of the BS, as shown in ref. \cite{Paris1996}. For this process we assume perfect overlap. The outcomes of this measurements in terms of the counts  $\{n_c,n_d\}$, are obtained through the APDs D1 and D2. It is important to mention that the displacement operation should also have spatio-temporal match with the coherent field used to build the HER, which have already been theoretically analyzed in noisy environments \cite{Wallentowitz1996}, and experimentally proven \cite{Lvovsky2002, Lvovsky2002a,Jeong2014}. When Bob measures, he also places a Gaussian spectral filter in front of his detectors, with central frequency and bandwidth $\omega_d$ and $\sigma_d$, respectively. This process is formally described as follows
		\begin{align}
		\rho^{' ABC}_{pd(pa)} &= \mathbb{I}_a \otimes \hat{\Pi}_b(n_b) \otimes \hat{\Pi}_c(n_c)\rho^{ABC}\mathbb{I}_a \otimes \hat{\Pi}_b(n_b) \\ \nonumber&\otimes \hat{\Pi}_c(n_c) / \text{Tr} [\mathbb{I}_a \otimes \hat{\Pi}_b(n_b) \otimes \hat{\Pi}_c(n_c)\rho^{ABC}].\label{abc}
		\end{align}
		We use primed density matrices for states after any measurement $\rho \to \rho '$. It is important to note that, the displaced number basis could be omitted if the HER were generated with a single-photon instead of a CS in BS1, i.e., for $\alpha=0$. This restriction would reduce the study to typical single-mode quantum teleportation setups, but still with one major difference: all the involved fields (and states) are realistic in the sense that no bandwidth restrictions are taken into account. Also, all the considered spectra are experimentally feasible. 
		
		\item Bob obtains the results for $\{n_c,n_d\}$ as $\{0,1\}$ or $\{1,0\}$, and communicates this result to Alice. 
		\item With Bob's results, Alice performs a unitary transformation and calculates an averaged version of the teleportation fidelity in the form
		
		\begin{equation}
		\begin{aligned}
		\label{eq:AvgFid}
		\overline{\mathcal{F}}_{pd(pa)}&= \frac{1}{\theta_f-\theta_i} \int\nolimits_{\theta_i}^{\theta_f} d\theta \text{Tr} [ \hat{\sigma}_i\rho^{' A}_{pd(pa)}(\theta)\hat{\sigma}_i^\dagger \\ &\times \hat{D}(\{\alpha\} )   \ket{\psi^C(\theta)} \bra{\psi^C (\theta)}\hat{D}^\dagger(\{\alpha\}) ],	
		\end{aligned}
		\end{equation}	
		
		where $\rho^{' A}_{pd(pa)}=\text{Tr}_{bc}\left[\rho^{' ABC}_{pd(pa)}\right]$ is the density matrix of Alice's post-selected state. The coefficients of the input states (the qubit) are considered real for simplicity and parametrized using $x=\sin (\theta)$, $y=\cos (\theta)$, $\hat{\sigma}_i$ is a Pauli matrix that conditionally depends on the outcomes $\{n_c,n_d\}$, and $\hat{D}(\{\alpha\} )$ stands for a displacement operator with $\alpha$ defined as in (\ref{eq:rhoPAPDSFWM}) this implies that the teleported state changes from the original basis to a displaced CV computational basis. In writing eq. (\ref{eq:AvgFid}) it was assumed that the state teleported to Alice is ideally of the form $\ket{\psi^{C'}}= x\KetC{\alpha} + y \hat{D}(\{\alpha\} ) \hat{c}^\dagger_\beta \ket{0}$.
	\end{enumerate}

	\section{Quantum teleportation in a specific situation}
	\begin{figure}
		\centering
		\includegraphics[width=9cm]{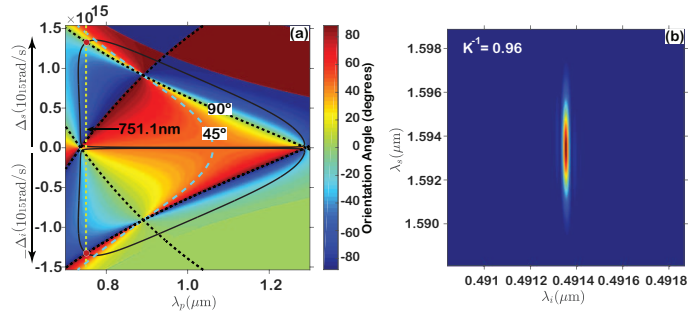}
		\caption{ (a) Phasematching contour (solid-black line) and phasematching function orientation angle (colored background), for FWM in the NKT photonics NL-PM-750 fiber, as function of the pump wavelength and emission frequencies, expressed in terms of detunings $\Delta_{s(i)}=\omega_{s(i)}-\omega_p$. (b) Joint spectral intensity function obtained for a pump centered at $751.1$ nm. For simulation, it were considered a fiber length $L=80$ cm, and a pump bandwidth 0.5nm.}
		\label{fig:1}
	\end{figure}
	In this section, we explore the application of the QT protocol outlined above. For HER generation, we consider that the FWM process takes place in a commercially nonlinear photonic crystal fiber (PCF), the NKT photonics NL-PM-750. The proposal focuses on a degenerate pump configuration, for which the phasematching condition is given by $\Delta k = 2k_p-k_s-k_i-\phi_{nl}=0$, with $\phi_{nl}$ the nonlinear phase shift due to self/cross phase modulation \cite{Sinclair2016,Garay-Palmett2007}. In order to herald a pure quantum state by detecting photons in the idler generation mode of the two-photon like quantum state in Eq.~(\ref{eq:JSAtotalPA}), the latter must be a factorable state. Factorable two-photon states generated by SFWM can be obtained if certain conditions on the group velocity of the interacting fields are fulfilled, technique known as group velocity matching (GVM) \cite{Garay-Palmett2007}. Besides, the separability of the two-photon state becomes achieved for specific combinations of fiber length and pump bandwidth. For these, GVM determines the orientation angle ($\theta_{si}$) of the joint spectral intensity function (JSI), which in the generation frequency space $\{\omega_s,\omega_i\}$ is given by 
	
	\begin{equation}
	\theta_{si}=-\arctan (\tau_s/\tau_i),
	\end{equation}
	where $\tau_\mu=L\left[   (\nu_g(\omega_p^0))^{-1} - (\nu_g(\omega_\mu^0))^{-1}     \right]$, $\nu_g(\omega_\mu^0)$ is the group velocity of the $\mu$ field evaluated at the central frequency $\omega_\mu^0$. As shown in reference \cite{Garay-Palmett2007}, factorable two-photon states can be obtained for $0\leq\theta_{si}\leq 90^\circ$.
	
\begin{figure}
		\centering
		\includegraphics[width=5cm]{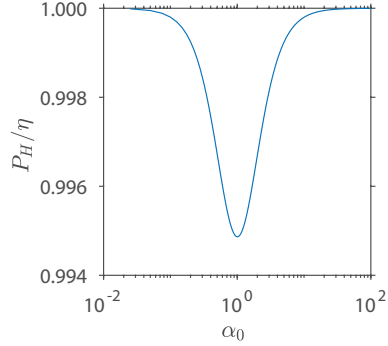}
		\caption{ Heralding efficiency $P_H / \eta$ of the SPS or SPACS as a function of $\alpha_0$.}
		\label{fig:PH}
	\end{figure}

	For the fiber considered in this proposal, solutions to the FWM phasematching condition ($\Delta k =0$) are shown as a solid black contour plot in Fig. \ref{fig:1}(a), where the horizontal and vertical axes correspond to pump frequencies and emission frequencies, respectively. Note that the latter are expressed in terms of detuning from the pump frequency. The corresponding orientation angle $\theta_{si}$ is represented in the same figure as the colored background, over which the contour plot for  $\theta_{si}=90^\circ$  and $\theta_{si}=45^\circ$ have been shown by the discontinuous black and cyan lines respectively. Intersections of the phasematching contour with a particular GVM contour determine the pump and emitted frequencies, for which a two-photon state with particular spectral correlation properties can be generated. Here, we are interested in the intersection that occurs at the pump wavelength $\lambda_p=751.1$nm, which is marked on the figure by red circles and correspond to a JSI orientation angle $\theta_{si}=90^\circ$. This case is of special interest, as a highly factorable state can be obtained with low to null spectral filtering \cite{Kaneda2016}. The resulting joint spectral intensity is shown in Fig. \ref{fig:1}(b), for which a typical full width at half maximum (FWHM) bandwidth of $0.5$nm for pico-second pulsed laser, and a fiber length of $L=80$cm were assumed. It is worth to point out that no spectral filtering has been considered in calculating the JSI, and even so the spectral distribution exhibits a factorable character, making it suitable for the generation of heralded quantum states described by Eq. (\ref{eq:heralded}). Of course this state will be a heralded single-photon state or SPACS depending on the value of $\alpha_0$. To make this clearer, we calculate the heralding efficiency of a SPACS defined as  $P_H = P_{AB}/P_{A}$, where $P_{AB}$ is the joint probability of detecting a SPS in the heralding arm and a SPACS in the heralded one, while $P_A$ is the probability of detecting a SPS in the heralding arm. $P_H$ probability is plotted in Fig. \ref{fig:PH}, assuming that both detections have an efficiency $\eta$. Note that $P_H$ is close to the heralding efficiency of a SPS ($ \alpha_0 =0$) for all values of $\alpha_0$.
	
	For HER preparation (see Eq. (\ref{eq:rhoPAPDSFWM})), we propose to combine the herald state generated under the conditions explained above with a coherent state. In search for the highest possible spatio-temporal overlap between these states, the CS could be generated by stimulated FWM in another PCF with the same dispersion characteristics as the one for generation of the heralded quantum states given in Eq. (\ref{eq:heralded}), and we assumed this matching for this specific situation. For simplicity, in the proposed setup detectors D1 and D2 (see figure \ref{fig:Esquema}(c)) are assumed to have perfect quantum efficiency, with a spectral response set to 10 nm FWHM and centered at the IR wavelength with the maximum emission probability from the SFWM process, which is $1.593\mu$m in the JSI function shown in Fig. \ref{fig:1}(b). As a first approach, we set the length of the heralded-state generation fiber to $L=80$cm, and vary the spectral properties involved in the qubit to teleport. 
	
\begin{figure}
		\centering
		\includegraphics[width=9cm]{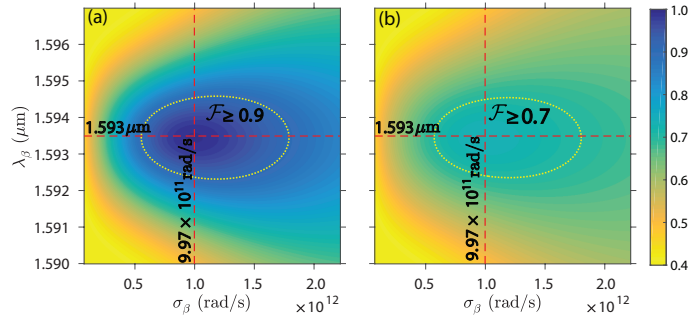}
		\caption{Average fidelities of the QT protocol dependent on the qubit central wavelength $\lambda_\beta$ and BW $\sigma_\beta$., using the HER: (a) $\rho_{pd}$ and (b) $\rho_{pa}$. Yellow-dotted line shows the spectral properties for a qubit to be teleported with high fidelity. The red-dotted lines intersect at the highest average fidelity for each set-up.}
		\label{fig:FidAvgd}
	\end{figure}

	Figure \ref{fig:FidAvgd} shows the average fidelity for both HERs $\rho_{pd}$ and $\rho_{pa}$, considering the parameters set as above. The fidelity is calculated from Eq. (\ref{eq:AvgFid}) in the $\theta$ domain $[0,2\pi]$, for a qubit with different Gaussian spectral envelope described by parameters $\{\lambda_\beta, \sigma_\beta  \}$, which are allow to vary. $\overline{\mathcal{F}}_{pd}$ is shown in 
	Fig. \ref{fig:FidAvgd}(a). It is interesting to point out that the value of $\overline{\mathcal{F}}_{pd}$ does not depend on $\alpha$. The dotted yellow line encloses the higher fidelities $\overline{\mathcal{F}}_{pd}\geq 0.9$ for $0.5639\leq\sigma_\beta\leq 1.789$ Trad/s. The red dotted lines indicate the values $\{\lambda_\beta, \sigma_\beta  \}$ (denoted as $\lambda_\beta^0=\lambda_\beta|_{\overline{\mathcal{F}}_{max}}$ and $\sigma_\beta^0=\sigma_\beta|_{\overline{\mathcal{F}}_{max}}$) that maximize the average fidelity in both the "pd" and "pa" and cases. On the other hand, $\overline{\mathcal{F}}_{pa}$ is shown in Fig. \ref{fig:FidAvgd}(b) with everything fixed as in the previous case, and the coherent state $\KetC{\alpha}$ described by a Gaussian spectral envelope with $\langle \hat{n} \rangle=0.5$ and equal spectral properties as the qubit, i.e., $\lambda_\alpha=\lambda_\beta$ and  $\sigma_\alpha=\sigma_\beta$ to maximize the temporal-modes overlap. The dotted yellow line encloses the higher fidelities $\overline{\mathcal{F}}_{pa}\geq 0.7$ for $0.5639\leq\sigma_\beta\leq 1.758$ Trad/s. The average fidelity obtained in the simulations show that even for the low value of $\langle \hat{n} \rangle=0.5$, the average fidelities for the "pa" case decrease substantially in comparison with the "pd" case. It is worth to point out the wide qubit-bandwidth interval for which the average quantum teleportation fidelity takes maximal values.
	
	
	It is well accepted that maximally entangled states are formed by pure wavefunctions \cite{Ekert95a,Eberly05a}; however, given the heralded nature of our sources to prepare the HER, this state is in general not pure but a statistical mixture as shown in eq. (\ref{eq:rhoPAPDSFWM}). However, the degree of "non-purity" is determined by the geometrical properties of the nonlinear medium \cite{Garay-Palmett2007}. Other decoherence processes, like photon loss or phase fluctuation would also imply a lower purity of the HER \cite{Dominguez-Serna2017}. To show the capabilities of this study, we have not included spectral filtering at any part of the heralding process of the SPS or SPACS needed to generate the HER. In addition to the average fidelity, we evaluate the HER capacity quantifying the purity of the SPS or the SPACS used to generate de HER, which can be obtained as the inverse of the cooperativity parameter $K$ given by \cite{URen2005}
	\begin{equation}
	K=\frac{1}{\sum_n \lambda_n^2},
	\end{equation}
	where $ \lambda_n$ are known as the Schmidt coefficients of $\ket{\psi_2}$ in eq. (\ref{eq:JSAtotalPA}), which can be obtained as the eigenvalues of the density matrix in eq. (\ref{eq:heralded}) for $\tilde{d}(\omega_i) =1$. The cooperativity parameter itself, gives information of the amount of entanglement of a bipartite state. This is calculated over the heralded resource alone, to show the influence of its purity on the entanglement usefulness as an information resource. 
	
\begin{figure}
		\centering
		\includegraphics[width=9cm]{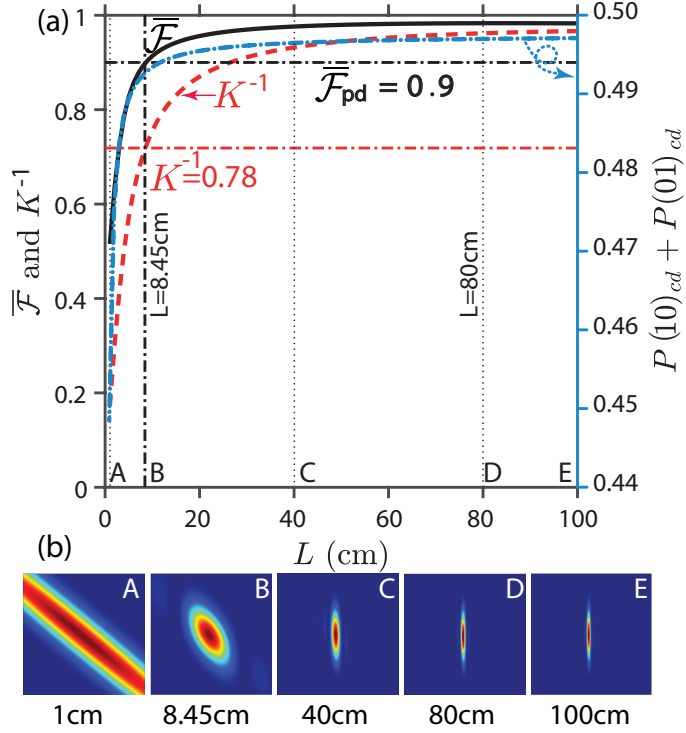}
		\caption{Fidelity and purity dependence on fiber lenght for the "pd" HER with the qubit spectral parameters $\lambda_\beta^0$ and $\sigma_\beta^0$ fixed. (a) Left axis: average fidelity $\overline{\mathcal{F}}_{pd}$ and purity $K^{-1}$ vs. $L$. Right axis: Probability of the joint measurement for the QT protocol vs. $L$; (b)Two-photon JSI for the corresponding fiber lengths marked as A to E.} 
		\label{fig:FidAvgdLdep}
	\end{figure}

	Fig. \ref{fig:FidAvgdLdep} shows the fidelity and purity dependence with fiber lenght for the "pd" HER with the qubit spectral parameters $\lambda_\beta^0$ and $\sigma_\beta^0$ fixed. This analysis is restricted to $\rho_{pd}$ given its higher fidelity when compared to the utilization of the HER $\rho_{pa}$. Fig. \ref{fig:FidAvgdLdep}(a) shows the average fidelity $\overline{\mathcal{F}}_{pd}$ and purity $K^{-1}$ on the left axis vs. $L$. Fidelities $\overline{\mathcal{F}}_{pd}\geq 0.9$ and purity $K^{-1}\geq0.7187$ are attainable for $L\geq 8.45$cm. While the probability of the joint measurement for the QT protocol is shown referred to the right axis, the nature of this Bell-like measurement sets an upper limit of 0.5 as only two outcomes are distinguishable \cite{Lutkenhaus1999}. In addition, \ref{fig:FidAvgdLdep}(b) shows the corresponding two-photon JSI for different fiber lengths from 1cm to 100cm as marked in panel (a) with vertical discontinuous lines. This is included for a better visualization of the influence of fiber length on the fidelity of the protocol and purity of the HER. We initially proposed $L=80$cm for this study, as we consider it a practical length to work with. It is important to note that even when high fidelities can be obtained with smaller lengths, the pair creation efficiency is proportional to $L$ \cite{Garay-Palmett2010}. It is clear from the figure that both $\overline{\mathcal{F}}_{pd}$ and $K^{-1}$ are close to $1$ for the chosen fiber length $L=80$cm.

	\section{Conclusions}
	We presented an experimentally feasible proposal for quantum teleportation based on a hybrid entangled state used as a resource that could be implemented  with available optical elements. Two entangled resources were studied, one formed by a heralded single photon combined with a coherent state in a beamsplitter and, the other combining a SPACS with the vacuum state in a BS. The capabilities of the proposal were proved by selecting a commercial PCF, that served to create the HER with a wide spectrum in the telecommunications band. High average fidelities $\overline{\mathcal{F}}_{pd} > 0.9$ are expected for the proposed "pd" HER.  Also the non-linear interaction length dependence of the average quantum teleportation fidelity as well as the purity of the heralded state was addressed. It was found that average fidelities higher than 0.9 are attainable even for purities lower than 0.8. Even when the qubit preparation was outside of the scope of this paper, it could easily be prepared by following a procedure similar to the one described in \cite{Babichev2004}. The present study can be straightforwardly applied to different geometries of the generation medium and other configurations on the degrees of freedom of the interacting fields, that for instance could involve polarization, spatial profiles, etc. Therefore, although the particular case studied shows an interesting flexibility on bandwidths of the teleported state, the finding of better candidates with higher flexibility is still an open research problem.
	
	\section{Funding Information}
	National Council of Science and Technology of Mexico, CONACyT (Catedras CONACyT) (709/2018).
	
	\section{Disclosures}
	The authors declare no conflicts of interest.

\end{document}